\documentstyle[aps]{revtex} \def\be{\begin{equation}} \def\ee{\end{equation}}
\def\bea{\begin{eqnarray}} \def\eea{\end{eqnarray}} \begin{document} \title {No
Quasi-long-range Order in Strongly Disordered Vortex Glasses:  a Rigorous Proof}

\author {D.E.  Feldman}

\address {Department of Condensed Matter Physics, Weizmann Institute of Science,
76100 Rehovot, Israel\\ and Landau Institute for Theoretical Physics, 142432
Chernogolovka, Moscow region, Russia} \maketitle

\begin{abstract} The paper contains a rigorous proof of the absence of
quasi-long-range order in the random-field $O(N)$ model for strong disorder in
the space of an arbitrary dimensionality.  This result implies that
quasi-long-range order inherent to the Bragg glass phase of the vortex system in
disordered superconductors is absent as the disorder or external magnetic field
is strong.  \end{abstract} 


The nature of the vortex phases of disordered superconductors is a subject of
active current investigations.  A plausible picture includes three phases
\cite{ph}:  vortex liquid (VL), vortex glass (VG) and Bragg glass (BG).  In all
those phases an Abrikosov lattice is absent \cite{Larkin} and only short-range
order (SRO) is expected in VG and VL.  A higher degree of ordering is predicted
in BG.  It is argued that in this state the vortex array is
quasi-long-range-ordered \cite{qlro1,qlro2}.  Thus, Bragg peaks can be observed
in BG \cite{Cubitt,yaron} as if the system had an Abrikosov lattice.  In the
other phases Bragg peaks are not found \cite{Cubitt}.  The phase transitions
from BG to VG and VL are presumably associated with topological defects
\cite{ph}.

The above picture is supported by variational and renormalization group
calculations for the random-field XY model \cite{qlro1,qlro2,var,var1} which is
the simplest model of the vortex array in disordered superconductors.  Besides,
this model is useful for our understanding of many other disordered systems
\cite{qlro2}.  Hence, the detailed knowledge of its properties is important.
Unfortunately, the only rigorous result about the random-field XY model is the
absence of long-range order (LRO) \cite{rig}.  The present paper contains a new
rigorous result:  QLRO is absent as the disorder is sufficiently strong.  It is
interesting that our proof is quite simple in contrast to the very nontrivial
demonstration of the absence of LRO for arbitrarily weak disorder \cite{rig}.

It is expected \cite{ph} that the vortex system of the disordered superconductor
has no ordering for strong disorder due to appearance of the dislocations.  The
XY model is a convenient polygon for understanding of the role of the
topological defects.  Since the vortex glass phase of disordered superconductors
corresponds to strong disorder, our rigorous result supports the conjecture that
the vortex glass phase has no QLRO.

The simplest model of the vortex array in disordered superconductors has the
following Hamiltonian \cite{qlro1,qlro2}

\be \label{1} H=\int d^3r [K(\nabla u({\bf r}))^2 + h\cos(2\pi u({\bf
r})/a-\theta({\bf r}))], \ee where $u$ is the vortex displacement, $a$ the
constant of the Abrikosov lattice in the absence of disorder, $\theta$ the
random phase.  The one-component displacement field $u({\bf r})$ describes
anisotropic superconductors.  The generalization for the isotropic case is
straightforward.  The ordering can be characterized in terms of the form-factor

\be \label{ff} G(r)=\langle\cos2\pi (u({\bf 0})-u({\bf r}))/a\rangle, \ee where
the angular brackets denote the thermal and disorder average.  This correlation
function contains information about neutron scattering.  LRO corresponds to a
finite large-distance asymptotics $G(r\rightarrow\infty)\rightarrow{\rm
constant}$, QLRO is described by the power law $G(r)\sim r^{-\eta}$ and SRO
corresponds to the exponential decay of the correlation function $G(r)$ at large
$r$.

We demonstrate that as the random-field amplitude $h$ Eq.  (\ref{1}) is large
the system possesses only SRO.  Since $h$ depends on the strength of the
disorder in the sample and the external magnetic field \cite{qlro2} we see that
SRO corresponds to the situation at which either the disorder or magnetic field
is strong.  We consider not only the XY model (\ref{1}) but also the more
general random-field $O(N)$ model.  Its Hamiltonian has the following structure

\be \label{2} H=-J\sum_{\langle ij\rangle} {\bf S}_i{\bf S}_j -H\sum_i {\bf
S}_i{\bf n}_i, \ee where ${\bf S}_i$ are the $p$-component unit spin vectors on
the simple cubic lattice in the D-dimensional space, ${\bf n}_i$ is the random
unit vector describing orientation of the random field at site $i$, the angular
brackets denote the summation over the nearest neighbors on the lattice.  The XY
Hamiltonian (\ref{1}) corresponds to $p=2$.  In this case the relation between
(\ref{1}) and (\ref{2}) is given by the formulae $S_x=\cos 2\pi u/a, S_y=\sin
2\pi u/a$, where $S_{x,y}$ are the spin components.

The idea of our proof is based on the fact that the orientation of any spin
${\bf S}_i$ depends mostly on the random fields at the nearest sites as the
random-field amplitude $H$ is sufficiently strong.  We shall see that the
knowledge of the random fields in the region of size $Nb$ with the center in
site $i$, where $b$ is the lattice constant, allows us to determine the
orientation of the spin ${\bf S}_i$ with the accuracy $\exp(-{\rm constant}N)$.
Thus, the orientations of any two distant spins depend on the realizations of
the random field in two non-intersecting regions up to exponentially small
corrections.  The values of the random field in these regions are uncorrelated.
Hence, the correlations of the distant spins are exponentially small.

Below we consider the case of the zero temperature.  Hence, the system is in the
ground state.  We assume that the amplitude of the random field

\be \label{3} H=2DJ(1+\sqrt{2}+\delta), \delta>0.  \ee Let us estimate the angle
between an arbitrary spin ${\bf S}_i$ and the local random field ${\bf
h}_i=H{\bf n}_i$.  The Weiss field ${\bf H}_W={\bf h}_i + {\bf H}_J$ acting on
the spin includes the random field ${\bf h}_i$ and the exchange contribution
${\bf H}_J=J\sum_j {\bf S}_j$, where ${\bf S}_j$ are the nearest neighbors of
the spin ${\bf S}_i$.  The latter $H_J\le 2DJ$ since the number of the nearest
neighbors is $2D$, where $D$ is the spatial dimension.  Hence, the minimal
possible modulus of the Weiss field is $(H-H_J)\ge (H-2DJ)$.  Any spin is
oriented along the local Weiss field ${\bf H}_W$.  Let us consider the triangle
two sides of which are ${\bf h}_i$ and ${\bf H}_J$, and the third side is
parallel to ${\bf S}_i$.  The laws of sinuses allow us to show that the maximal
possible angle between ${\bf h}_i$ and ${\bf S}_i$ is $\phi_1=\arcsin(2DJ/H)$.

We shall now determine the orientation of the spin ${\bf S}_i$ iteratively.  Let
the zero approximation ${\bf S}^0_i$ be oriented along the random field ${\bf
h}_i$.  Let the first approximation ${\bf S}^1_i$ be oriented along the Weiss
field ${\bf H}_W^0$, calculated with the zero approximation for the neighboring
spins:  ${\bf H}_W^0={\bf h}_i+J\sum_j {\bf n}_j$, where $\sum_j$ denotes the
summation over the nearest neighbors.  The second approximation ${\bf S}^2_i$ is
determined with the Weiss field in the first approximation, etc.  Any
approximation ${\bf S}^k_i$ depends on the random fields only at a finite set of
the lattice sites.  The distance between any such site and site $i$ is no more
than $kb$, where $b$ is the lattice constant.

In any approximation the Weiss field $H_W^k\ge H-2DJ$.  Let ${\bf s}_i^k$ be the
difference between the $k$th and $(k-1)$th approximations for the spin ${\bf
S}_i$.  Then $|{\bf H}_W^k-{\bf H}_W^{(k-1)}|\le 2DJm^k$, where $m^k$ denotes
the maximal value of $s_l^k$.  Hence, we find with the laws of sinuses that the
angle between ${\bf S}_i^k$ and ${\bf S}_i^{(k-1)}$ is less than
$\phi_k=\arcsin(2DJm^k/(H-2DJ))$.  Since $s^{(k+1)}_i\le 2\sin(\phi_k/2)$, one
obtains the following estimation:

\bea \label{4} m^{(k+1)}\le \sqrt{2\{1-\sqrt{1-[2DJm^k/(H-2DJ)]^2}\}}=
\sqrt{2[2DJm^k/(H-2DJ)]^2/[1+\sqrt{1-[2DJm^k/(H-2DJ)]^2}]}\le & & \nonumber \\
2\sqrt{2} D J m^k / (H-2DJ) = m^k/[1+\delta/\sqrt{2}]\le
m^1/[1+\delta/\sqrt{2}]^{k}, & & \eea where Eq.  (\ref{3}) is used.

Now we are in the position to estimate the correlation function (\ref{ff}).  In
terms of the $O(N)$ model it is given by the expression

\be \label{5} G(r)=\langle{\bf S}({\bf 0}){\bf S}({\bf r})\rangle.  \ee Let
$N=[r/2b]-1$, where the square brackets denote the integer part.  We decompose
the values of the spins in the following way ${\bf S}({\bf x})={\bf S}^N({\bf
x})+\sum_{k>N} {\bf s}^k({\bf x})$.  The $N$th approximations ${\bf S}^N({\bf
x})$ and ${\bf S}^N({\bf 0})$ depend on the orientations of the random fields in
different regions which have no intersection.  Hence, the correlation function
$\langle {\bf S}^N({\bf 0}){\bf S}^N({\bf r})\rangle$ is the product of the
averages of the two multipliers ${\bf S}^N$ and thus equals to zero due to the
isotropy of the distribution of the random field.  All other contributions to
Eq.  (\ref{5}) can be estimated with Eq.  (\ref{4}) and are exponentially small
as the functions of $N$.  This proves that the correlation function $G(r)$ has
an exponentially small asymptotics at large $r$.  Thus, for strong disorder
(\ref{3}) both LRO and QLRO are absent.

A challenging question concerns the presence of QLRO in the weakly disordered
random-field systems.  Another interesting related problem is the question about
QLRO in the random-anisotropy $O(N)$ model \cite{F}.  Unfortunately, our
approach cannot be directly generalized for this problem since the zeroth-order
approximation ${\bf S}^0_i$ is not unique in the random-anisotropy model:  any
spin has two preferable orientations.

In conclusion, we have proved that for strong disorder the random-field $O(N)$
model has no QLRO.  The renormalization group calculations \cite{F} suggest that
at $N>2$ QLRO is absent for arbitrarily weak disorder.  On the other hand, in
the random-field $O(2)$ model without dislocations the renormalization group
\cite{qlro2} predicts QLRO.  Our result shows that in the system with the
topological defects QLRO is absent at least as the disorder is strong.  This
prediction is relevant for the vortex glass state of the disordered
superconductors.

 \end{document}